\documentclass[5p]{elsarticle}

\usepackage{amsmath}
\usepackage{isotope}
\usepackage{flushend}

\biboptions{sort&compress}

\usepackage[normalem]{ulem}  
\usepackage[dvips]{color} 

\renewcommand\sout{\bgroup \color{red} \ULdepth=-.5ex \ULset}

\begin{document}

\title{Constraining the in-medium nucleon-nucleon cross section from the width of nuclear giant dipole resonance}


\author[1,2]{Rui Wang}
\ead{wangrui@sinap.ac.cn}

\author[3]{Zhen Zhang}
\ead{zhangzh275@mail.sysu.edu.cn}

\author[4]{Lie-Wen Chen}
\ead{lwchen@sjtu.edu.cn}

\author[5]{Che Ming Ko}
\ead{ko@comp.tamu.edu}

\author[1,2]{Yu-Gang Ma}
\ead{mayugang@fudan.edu.cn}

\address[1]{Key Laboratory of Nuclear Physics and Ion-beam Application~(MOE), Institute of Modern Physics, Fudan University, Shanghai $200433$, China}

\address[2]{Shanghai Institute of Applied Physics, Chinese Academy of Sciences, Shanghai $201800$, China}

\address[3]{Sino-French Institute of Nuclear Engineering and Technology,
Sun Yat-Sen University, Zhuhai $519082$, China}

\address[4]{School of Physics and Astronomy and Shanghai Key Laboratory for Particle Physics and Cosmology, Shanghai Jiao Tong University, Shanghai $200240$, China}

\address[5]{Cyclotron Institute and Department of Physics and Astronomy, Texas A$\&$M University, College Station, Texas $77843$, USA}


\begin{abstract}
We develop a new lattice Hamiltonian method for solving the Boltzmann-Uehling-Uhlenbeck~(BUU) equation. Adopting the stochastic approach to treat the collision term and using the GPU parallel computing to carry out the calculations allows for a rather high accuracy in evaluating the collision term, especially its Pauli blocking, leading thus to a new level of precision in solving the BUU equation. Applying this lattice BUU method to study the width of giant dipole resonance~(GDR) in nuclei, where the accurate treatment of the collision term is crucial, we find that the obtained GDR width of \isotope[208]{Pb} shows a strong dependence on the in-medium nucleon-nucleon cross section $\sigma_{\rm NN}^*$. A very large medium reduction of $\sigma_{\rm NN}^*$ is needed to reproduce the measured value of the GDR width of \isotope[208]{Pb} at the Research Center for Nuclear Physics in Osaka, Japan.
\end{abstract}


\maketitle

\emph{Introduction.}
The in-medium nucleon-nucleon~(NN) cross section $\sigma_{\rm NN}^*$ has significant effects on the dynamics of heavy-ion collisions (HICs), and it thus plays a crucial role in understanding the reaction mechanisms as well as various phenomena and observables in these collisions~\cite{WesPRL71,CXPRC58,CLWPLB459,LJYPRL86,WTTPRC97,LPCNST29}.  The importance of $\sigma_{\rm NN}^*$ also lies in its intimate relation to the transport properties of nuclear matter~\cite{LHLPRC96,BarPRC99} and the nucleon effective interactions~\cite{SamPRC73}. Since a major goal of studying HICs is to extract the equation of state~(EOS) of nuclear matter from experimental data~\cite{LBAPRL78,Dansc298,CLWPRL94,FamPRL97,TsaPRL102,OLPRL115}, a thorough understanding of $\sigma_{\rm NN}^*$ helps reduce the uncertainties in transport models~\cite{XJPRC93,ZYXPRC97} that are used for describing these reactions.  While the NN cross section in free space $\sigma_{\rm NN}^{\rm free}$ can be directly measured in experiments, the determination of the value of $\sigma_{\rm NN}^*$ in nuclear medium usually relies on theoretical investigations. These include calculations based on microscopic theories, such as the nonrelativistic and relativistic Brueckner theories~\cite{LGQPRC48,LGQPRC49,AlmPRC50,SchPRC55,KohPRC57,FucPRC64,SamPRC73} and the closed time path Green's function approach~\cite{MGPRC49,LQPRC69}.
Also, there have been attempts to extract $\sigma_{\rm NN}^*$ from experiments by comparing results from transport model calculations, where $\sigma_{\rm NN}^*$ is a crucial input, with observables measured in HICs that are sensitive to $\sigma_{\rm NN}^*$, e.g., the collective flow and nuclear stopping~\cite{WesPRL71,LBAPRC71,LBAPRC72,ZYPRC75,LopPRC90,BarPRC99,OLCPC43}. Although these studies have reached the consensus that the NN cross section is suppressed in nuclear medium, the reduction factor is still far from certainty.

The transport model used in describing HICs is a straightforward tool for studying $\sigma_{\rm NN}^*$ because one of its main ingredients, the NN collision term, embodies the information of $\sigma_{\rm NN}^*$.
Since the mean field or the EOS of nuclear matter is another major ingredient of one-body transport models, finding the proper observables that depend on $\sigma_{\rm NN}^*$ rather than the nuclear EOS is essential for studying $\sigma_{\rm NN}^*$. One such observable is the width of nuclear giant dipole resonance~(GDR) as it is naturally related to $\sigma_{\rm NN}^*$ through the NN collision term in transport models. In general, the damping width of nuclear collective motion originates from three sources: 1)~the escape width associated with particle emissions;  2)~the fragmentation or the Landau damping width due to couplings between single particle states and the mean field;  3)~the spreading or collisional damping width caused by the coupling to more complex states like the two-particle-two-hole~($2p$-$2h$), $3p$-$3h$, etc. For a heavy nucleus at zero temperature, the width of its GDR is mainly exhausted by collisional damping~\cite{KolPRC54,DanPRC84,GarPPNP101} before the contribution from deformation fluctuations appears as a result of the finite temperature effect~\cite{BraPRL74}. In the transport model, the collisional damping is incorporated in the binary collisions of nucleons and thus depends directly on $\sigma_{\rm NN}^*$. It is therefore expected that the GDR width of a heavy nucleus in studies based on the transport model depends strongly on $\sigma_{\rm NN}^*$ and weakly on the nuclear EOS.

The major obstacle that has so far prevented the use of transport models to accurately calculate the spreading width of GDR is due to the fermionic nature of nucleons. Specifically, the accurate treatment of Pauli blocking in transport models is challenging~\cite{XJPRC93,ZYXPRC97}, especially for small amplitude nuclear collective motions. Both subtle implementations and advanced computing techniques are required for overcoming this difficulty. In the present work, we extend the previous study using the lattice Hamiltonian Vlasov method based on the next-to-next-to-next-to leading order~(N$3$LO) Skyrme pseudopotential~\cite{WRPRC99} to include a stochastic elastic NN collision term. Solving the resulting Boltzmann-Uehling-Uhlenbeck~(BUU)-type one-body transport model with the high computation efficiency provided by GPU parallel computing~\cite{Rue2013}, which enables the accurate treatment of Pauli blocking in the collision term of the BUU equation, allows us to calculate precisely the spreading width of the GDR in nuclei. We then obtain a stringent constraint on the in-medium NN cross section $\sigma_{\rm NN}^*$ by comparing the GDR width of \isotope[208]{Pb} from the present lattice BUU~(LBUU) method with that measured from $\isotope[208]{Pb}(\vec{p},\vec{p}')$ reaction with polarized protons at the Research Center for Nuclear Physics~(RCNP) in Osaka, Japan~\cite{TamPRL107}.

\emph{Model description.}
The BUU equation is a semi-classical approximation to the quantum transport equation~\cite{CarRMP55,BerPR160}.
For a momentum-dependent mean-field potential $U(\vec{r},\vec{p})$, it reads as
\begin{equation}\label{E:BUU}
    \frac{\partial f}{\partial t} + \frac{\vec{p}}{E}\cdot\nabla_{\vec{r}}f + \nabla_{\vec{p}}U(\vec{r},\vec{p})\cdot\nabla_{\vec{r}}f - \nabla_{\vec{r}}U(\vec{r},\vec{p})\cdot\nabla_{\vec{p}}f = I_{\rm c},
\end{equation}
where $f=f(\vec{r},\vec{p})$ is the one-body phase-space distribution function of nucleons or their Wigner function.
The r.h.s of Eq.~(\ref{E:BUU}) is the NN collision term including the Pauli blocking effect due to the Fermi statistics of nucleons, i.e.,
\begin{equation}
\begin{split}
    I_{\rm c} = &-g\int\frac{{\rm d}\vec{p}_2}{(2\pi\hbar)^3}\frac{{\rm d}\vec{p}_3}{(2\pi\hbar)^3}\frac{{\rm d}\vec{p}_4}{(2\pi\hbar)^3}|{\cal M}_{12\rightarrow34}|^2\\
    &\times(2\pi)^4\delta^4(p + p_2 - p_3 - p_4)\\
    &\times[ff_2(1-f_3)(1-f_4) - f_3f_4(1 - f)(1 - f_2)],
\end{split}\label{E:Ic}
\end{equation}
where $g$ is the degeneracy, ${\cal M}_{12\rightarrow34}$ is the in-medium transition matrix element, and $(1-f_i)$ is the Pauli suppression factor. It is worth mentioning that higher-order quantum corrections to Eq.~(\ref{E:BUU}) can be added perturbatively~\cite{BonPRL71}.

In the present work, we solve the BUU equation by the lattice Hamiltonian~(LH) method~\cite{LenPRC39,XHMPRL65,XHMPRL67}, which is a variant of the usual test particle method~\cite{WonPRC25}. In the LH method, the total Hamiltonian $H$ of the system is approximated by the lattice Hamiltonian $H_L$, i.e., 
\begin{equation}\label{E:HL}
    H = \int {\cal H}(\vec{r}){\rm d}\vec{r} \approx l^3\sum_{\alpha}{\cal H}(\vec{r}_{\alpha})\equiv H_L,
\end{equation}
where $\mathcal{H}$ is the Hamiltonian density, $\vec{r}_{\alpha}$ represents the coordinate of certain lattice site $\alpha$, and $l$ is the lattice spacing. For the nucleon one-body phase-space distribution function $f_{\tau}(\vec{r}_{\alpha},\vec{p})$, it is expressed as
\begin{equation}\label{E:f}
    f_{\tau}(\vec{r}_{\alpha},\vec{p},t) = \frac{(2\pi\hbar)^3}{gN_{\rm E}}\sum_i^{\alpha,\tau}S\big[\vec{r}_i(t) - \vec{r}_{\alpha}\big]\delta\big[\vec{p}_i(t) - \vec{p}\big],
\end{equation}
where $S$ is the form factor and $N_{\rm E}$ is the number of ensembles or test particles used in the calculation. The sum in Eq.(\ref{E:f}) runs over all test nucleons of isospin state $\tau$ that contribute to the lattice site $\alpha$.
In the present work, we adopt a triangular form factor $S$ with the size of $4l$, and its detail can be found in Ref.~\cite{WRPRC99}.
The Hamiltonian in Eq.~(\ref{E:HL}) contains both the Coulomb and the nuclear part~\cite{WRPRC99} with the latter obtained from the N$3$LO Skyrme pseudopotential~\cite{RaiPRC83} SP$6$h, whose details can be found in Ref.~\cite{WRPRC98}.

In the present LBUU method, the ground state of a spherical nucleus at zero-temperature is obtained from the Thomas-Fermi approach~\cite{LenPRC39,DanNPA673,GaiPRC81,LHPRC99} via the variation of the Hamiltonian with respect to the radial nucleon density $\rho_{\tau}(r)$.  The obtained $\rho_{\tau}(r)$ is then used to determine the initial coordinates of test nucleons, while their initial momenta are generated according to zero-temperature Fermi distribution with local Fermi momentum given by $p_{\tau}^{F}(r)$ $=$ $\hbar\big[3\pi^2\rho_{\tau}(r)\big]^{1/3}$.
This method for initialization ensures the stability of ground-state nuclei in BUU-like transport models~\cite{LHPRC99,WRPRC99}. 

For the collision term in the BUU equation, we implement it using the stochastic approach~\cite{DanNPA533}, which is more reliable than the commonly used geometric method when the mean free path $\lambda_{\rm MFP}$ of a test nucleon is not much larger than the interaction length between two test nucleons~\cite{XZPRC71} or when the NN scattering cross section is very large. The collision probability $P_{ij}$ of two test nucleons in the stochastic approach is determined from the NN collision term in Eq.~(\ref{E:Ic}), which is
\begin{equation}
    P_{ij} = v_{\rm rel}\sigma_{\rm NN}^*S(\vec{r}_i - \vec{r}_{\alpha})S(\vec{r}_j - \vec{r}_{\alpha})l^3\Delta t.
\end{equation}

To reduce the statistical fluctuations of collision events and better reflect the nature of the BUU equation, we include collisions of test nucleons from different ensembles.  In this case, the collision probability is reduced to $P_{ij}/N_{\rm E}$, because of the scaling $\sigma_{\rm NN}^*$ $\rightarrow$ $\sigma_{\rm NN}^*/N_{\rm E}$ of the in-medium NN cross section between test nucleons.  Under such a scaling, the diluteness of the system, which is characterized by $\sqrt{\sigma_{\rm NN}^*}/\lambda_{\rm MFP}$, is reduced by the factor $\sqrt{N_{\rm E}}$, and this makes it possible to solve the BUU equation almost exactly with a sufficiently large $N_{\rm E}$ achieved by adopting the GPU parallel computing.

For the $i$-th and $j$-th test nucleons colliding at the lattice site $\vec{r}_{\alpha}$, the direction of their final momenta $\vec{p}_3$ and $\vec{p}_4$ are sampled according to the differential cross-section given in Ref.~\cite{CugNIMB111}.
However, this collision can only happen if it is allowed by the Pauli principle via the factor $[1 - f(\vec{r}_{\alpha},\vec{p}_3)]\times[1 - f(\vec{r}_{\alpha},\vec{p}_4)]$.  In the present LBUU method, the distribution function $f_{\tau}(\vec{r}_{\alpha},\vec{p})$ is calculated from averaging its value in Eq.~(\ref{E:f}) over a given momentum-space sphere centered at $\vec{p}$ with radius $R_{\tau}^p(\vec{r}_{\alpha},\vec{p})$.
In typical transport model calculations, $R_{\tau}^p(\vec{r}_{\alpha},\vec{p})$ is taken to have a constant value of about one hundred MeV.  In the present work, we use an improved form for $R_{\tau}^p(\vec{r}_{\alpha},\vec{p})$ that is specifically proposed for small-amplitude nuclear collective dynamics near ground state~\cite{GaiPRC81}, i.e., $R_{\tau}^p(\vec{r}_{\alpha},\vec{p})$ $=$ ${\rm max}[\Delta p, p_{\tau}^F(\vec{r}_{\alpha}) - |\vec{p}|]$,
where $p_{\tau}^F=\hbar(3\pi^2\rho_{\tau})^{1/3}$ is the local nucleon Fermi momentum and $\Delta p$ is a constant with the dimension of momentum that needs to be taken to be sufficiently small.

The treatment of Pauli blocking in transport models is crucial in calculating the width of nuclear collective excitations. At low incident energy or temperature, the Pauli blocking is notoriously difficult to handle in transport models~\cite{XJPRC93,ZYXPRC97}. This is mainly caused by the inaccuracy in calculating the local momentum distribution $f_{\tau}(\vec{r}_{\alpha},\vec{p})$, which then leads to numerically spurious collisions and thus an overestimated GDR width as a result of the enhanced collisional damping. There are three main origins for the numerically spurious collisions in transport models: 
1) fluctuations in calculating $f_{\tau}(\vec{r}_{\alpha},\vec{p})$ caused by insufficiently large $N_{\rm E}$; 2) spurious thermal excitation caused by finite $\Delta p$ in calculating $f_{\tau}(\vec{r}_{\alpha},\vec{p})$~(also see Ref.~\cite{GaiPRC81}); and 3) diffusion in local momentum caused by finite lattice spacing $l$ when averaging over different local densities on the nuclear surface.

In choosing the parameter values in the LBUU calculations, we use the following criteria. For a given $l$ and $\Delta p$, $N_{\rm E}$ should be large enough to eliminate the overwhelming majority of the spurious collisions caused by the first origin mentioned above, and at the same time $l$ and $\Delta p$ should be chosen to be sufficiently small to suppress the effects due to the second and third origins on the GDR width. After careful tests based on considerations of numerical accuracy and computation efficiency, we find the optional values of $l$ $=$ $0.5~\rm fm$, $\Delta p$ $=$ $0.05~\rm GeV$ and $N_{\rm E}$ $=$ $30000$. It is worth to mention that with the adoption of GPU parallel computing, it is possible to use a value for $N_{\rm E}$ that far exceeds those used in all previous calculations based on the BUU transport equation. Further reducing $\Delta p$ and $l$ and increasing $N_{\rm E}$ only leads to a negligible variation in the calculated GDR width.

We note that for the case of free NN cross section, an average of $97.93\%$ of the attempted collisions in the ground state of \isotope[208]{Pb} are blocked by the Pauli principle, resulting in an average of $1.30$ successful collisions of physical nucleons per ${\rm fm}/c$ during the time evolution of $0-500$ ${\rm fm}/c$.  Also, the root mean square (rms) radius and the ground-state energy of \isotope[208]{Pb} vary by less than $3.6\%$ ($0.2~\rm fm$) and $3.2\%$ ($50~\rm MeV$), respectively, during this time evolution. With a reduced in-medium NN cross section, both the number of successful collisions and the change in the radius and binding energy are even smaller.
Since the binding energy decreases monotonically with time without oscillations, it is not expected to have much effect on the calculated excitation energy of GDR.
The energy violation, which is caused by our use of in-vacuum energy conservation in NN scatterings, instead of the in-medium energy conservation in the presence of the momentum-dependent potential, is not expected to affect the calculated width of GDR either.
This is because the latter is controlled by the NN scattering rate, which depends on the NN scattering cross section and the Pauli blocking factor.
We also note that although the radius and binding energy variations in LBUU are larger than those in the lattice Hamiltonian Vlasov approach of Ref.~\cite{WRPRC99}, where the rms radius and the binding energy almost do not change, and the radial density profile only changes slightly during the time evolution of 0-1000 fm/c, the LBUU method used in the present study is sufficiently accurate for investigating the GDR width.

\emph{Results and discussions.}
The collective excitation of a nucleus consisting of $A$ nucleons can be induced by adding a perturbation to its Hamiltonian at the initial time $t_0$, i.e., $\hat{H}_{ex}(t)$ $=$ $\lambda\hat{Q}\delta(t - t_0)$, where $\hat{Q}$ is an appropriate excitation operator and $\lambda$ is a small parameter. The width of a collective excitation is defined by the full width at half maximum~(FWHM) of its strength function $S(E)$ as a function of the excitation energy $E$.  In the linear response theory~\cite{Fet1971}, the $S(E)$ is obtained from the Fourier integral
\begin{equation}\label{E:S-Q}
    S(E) = -\frac{1}{\pi\lambda}\int_0^{\infty}{\rm d}t\Delta\langle\hat{Q}\rangle(t){\rm sin}\frac{Et}{\hbar},
\end{equation}
where $\Delta\langle\hat{Q}\rangle(t)$ $=$ $\langle 0'|\hat{Q}|0'\rangle$ $-$ $\langle0|\hat{Q}|0\rangle$ is the time evolution of the response function of the nucleus to the excitation operator $\hat{Q}$
with $|0\rangle$ and $|0'\rangle$ denoting the nuclear states before and after the perturbation, respectively.  In terms of the Wigner transform $q(\vec{r},\vec{p})$ of the one-body excitation operator $\hat{q}$, which is related to $\hat Q$ by $\hat{Q}$ $=$ $\sum_i^{A}\hat{q}$, the expectation values in the above can be evaluated according to $\langle\hat{Q}\rangle(t) = \int f(\vec{r},\vec{p},t)q(\vec{r},\vec{p}){\rm d}\vec{r}{\rm d}\vec{p}$ using the nucleon phase-space distribution function $f(\vec{r},\vec{p},t)$.
Details on the single-particle operator used in exciting a ground state nucleus in transport models can be found in Ref.~\cite{WRPRC99}.

\begin{figure}[!hbt]
\centering
\includegraphics[width=8.5cm]{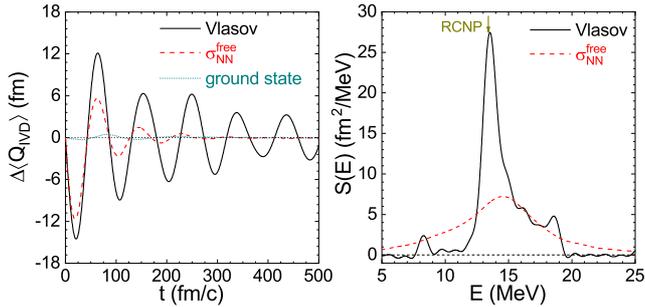}
\caption{Time evolution of the isovector dipole response function $\Delta\langle\hat{Q}_{\rm IVD}\rangle$~(left) and strength function $S(E)$~(right) of \isotope[208]{Pb} due to the perturbation of $\hat{H}_{ex}$ $=$ $\lambda\hat{Q}_{\rm IVD}\delta(t - t_0)$ with $\lambda$ $=$ $15~ {\rm MeV}/c$ from the Vlasov calculation (solid lines) and the LBUU calculation (dashed lines) with $\sigma_{\rm NN}^{\rm free}$.
The dotted cyan curve in the left window represents the expectation value of the $\hat{Q}_{\rm IVD}$ in the ground state of \isotope[208]{Pb} from the LBUU calculation with $\sigma_{\rm NN}^{\rm free}$.}
\label{F:QDV}
\end{figure}

We first employ the present LBUU method to study the effect of NN scatterings on the isovector dipole response of \isotope[208]{Pb} using the excitation operator $\hat Q_{\rm IVD}$ $=$ $\frac{N}{A}\sum_i^Z\hat z_i$ $-$ $\frac{Z}{A}\sum_i^N\hat z_i$.
In Fig.~\ref{F:QDV}, we show the results obtained by using the free NN elastic scattering cross section taken from Ref.~\cite{CugNIMB111} with $\sigma_{\rm NN}^{\rm free}(p_{\rm lab}) = \sigma_{\rm NN}^{\rm free}(0.1~{\rm GeV}/c)$ for neutron-neutron~($nn$) or proton-proton~($pp$) collisions at $p_{\rm lab}\leq 0.1~{\rm GeV}/c$ and $\sigma_{\rm NN}^{\rm free}(p_{\rm lab}) = \sigma_{\rm NN}^{\rm free}(0.05~{\rm GeV}/c)$ for neutron-proton~($np$) collisions at $p_{\rm lab}\leq 0.05~{\rm GeV}/c$, as experimental data for lower incident momenta~($p_{\rm lab}$) are unavailable. For comparison purpose, results from the LBUU calculation without NN scatterings, i.e., the Vlasov calculation, are also shown in Fig.~\ref{F:QDV}. In both cases, we use in the initial perturbation the same parameter $\lambda$ $=$ $15~{\rm MeV}/c$, which is also used in all the calculations in the present study, and we find that varying the value of $\lambda$ by $2/3$ almost has no effects on the value of the GDR width. As shown in the left window of Fig.~\ref{F:QDV} for the response function $\Delta\langle\hat{Q}_{\rm IVD}\rangle(t)$, the inclusion of NN scatterings significantly enhances the damping of the oscillations.
The dotted cyan curve in the left window of Fig.~\ref{F:QDV} represents the expectation value of the $\hat{Q}_{\rm IVD}$ in the ground state of \isotope[208]{Pb} from the LBUU calculation with $\sigma_{\rm NN}^{\rm free}$, which is negligible compared with that in the excited cases. To illustrate more clearly the effect of collisional damping, we show in the right window of Fig.~\ref{F:QDV} the GDR strength function $S(E)$ from the Fourier transformation of the response function. Note that the Vlasov calculation is carried out for a long evolution time of $1000~{\rm fm}/c$ when the amplitude of the oscillation of $\Delta\langle\hat{Q}\rangle(t)$ almost vanishes so that the fluctuation in the calculated strength function  from the Fourier transform of $\Delta\langle\hat{Q}\rangle(t)$ is negligible. We clearly see the large increase of GDR width after including NN scatterings, namely, the GDR width of $^{208}$Pb are $6.5~\rm MeV$ and $1.5~\rm MeV$ in the LBUU calculations with and without NN scatterings, respectively. 

\begin{figure}[!hbt]
\centering
\includegraphics[width=8cm]{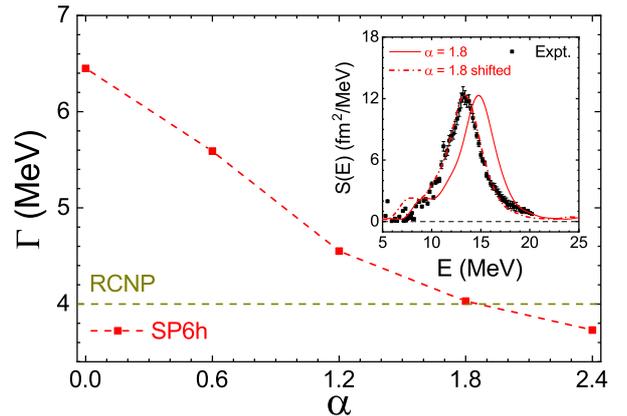}
\caption{The GDR width of \isotope[208]{Pb} from LBUU calculations for different values of $\alpha$ in $\sigma_{\rm NN}^*$. The horizontal line represents the RCNP experimental value of $4.0~\rm MeV$. The inset shows the strength function with $\alpha$ $=$ $1.8$~(solid line) and the shifted one~(dash-dotted line) to match the experimental GDR peak energy. See text for details.}
\label{F:gm}
\end{figure}

Experimentally, the GDR width of \isotope[208]{Pb} has been well determined to be $4.0~\rm MeV$ from the $\isotope[208]{Pb}(\vec{p},\vec{p}')$ reaction carried out at RCNP~\cite{TamPRL107}. Our result from the LBUU calculation with $\sigma_{\rm NN}^{\rm free}$ thus significantly overestimates the GDR width of \isotope[208]{Pb}. This is understandable because of the absence of medium effect on the NN scattering in the calculation.  Its inclusion is expected to reduce the NN cross section, weaken the collisional damping, and result in a smaller GDR width. The sensitivity of the GDR width to NN scatterings shown in Fig.~\ref{F:QDV} makes it possible to constrain the medium effect on the NN scattering cross section.

For $\sigma_{\rm NN}^*$, we parameterize it by multiplying the free NN cross section with a medium-dependent correction factor. Specifically, we choose an exponential reduction factor as suggested by the $T$-matrix approach in Ref.~\cite{AlmPRC50}, i.e.,
\begin{equation}\label{E:ECS}
    \sigma_{\rm NN}^* = \sigma_{\rm NN}^{\rm free}{\rm exp}\bigg[-\alpha\frac{\rho/\rho_{\rm nuc}}{1 + (T_{\rm c.m.}/T_0)^2}\bigg].
\end{equation}
In the above, $T_{\rm c.m.}$ is the the total kinetic energy of two scattering test nucleons at the rest frame of the local medium or cell, $\rho_{\rm nuc}$ $=$ $0.16~{\rm fm}^{-3}$ is the nuclear normal density, and $T_0$ $=$ $0.015~{\rm GeV}$. For the parameter $\alpha$, its original value in Ref.~\cite{AlmPRC50} is $0.6$, which is called the Rostock cross section. In the present study, we treat it as a free parameter to control the strength of medium effect. Displayed in Fig.~\ref{F:gm} is the GDR width $\Gamma$ of \isotope[208]{Pb} obtained with different values of $\alpha$. As expected, the GDR width decreases with increasing $\alpha$.  To reproduce the experimental value of $\Gamma$ $=$ $4.0~{\rm MeV}$ measured at RCNP requires $\alpha$ to be as large as about $1.8$, which indicates a very large medium reduction of the NN scattering cross section.

\begin{figure}[!hbt]
\centering
\includegraphics[width=8.5cm]{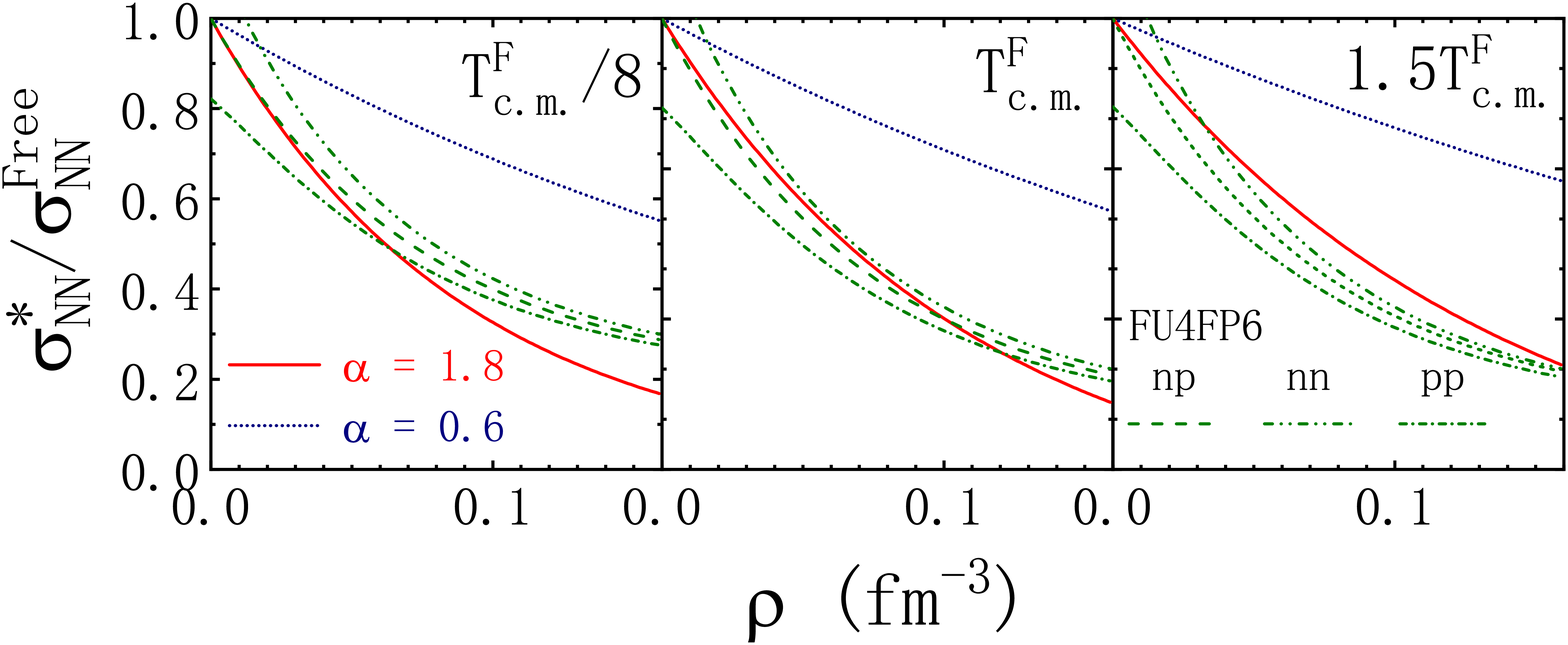}
\caption{Density dependence of the medium correction at different values for the total kinetic energy $T_{\rm c.m.}$ of two scattering nucleons using NN cross sections from Eq.~(\ref{E:ECS}) with $\alpha$ $=$ $1.8$, the Rostock cross section with $\alpha$ $=$ $0.6$, and the FU$4$FP$6$ parameterization.}
\label{F:MC}
\end{figure}

Although an early study on the balance energy, at which the nucleon direct flow in HICs vanishes,  favors a small medium reduction of the NN cross section~\cite{WesPRL71}, more recent studies based on the analysis of the collective flow and nuclear stopping data~\cite{LPCNST29,LopPRC90} as well as the nucleon induced reaction cross section~\cite{OLCPC43} require a large medium reduction. For comparisons, we also calculate the GDR width from the LBUU method with two different $\sigma_{\rm NN}^*$, namely, the Fuchs cross section~\cite{FucPRC64}, which is obtained from the in-medium Dirac-Brueckner T matrix, and the FU$4$FP$6$ parameterization, which is preferred by the nucleon induced reaction cross section~\cite{OLCPC43}. The value of GDR width of \isotope[208]{Pb} calculated using the FU$4$FP$6$ parameterization is $4.32~\rm{MeV}$, which is consistent with the experimental data. On the other hand, the values obtained with the Fuchs cross section and the Rostock cross section with $\alpha$ $=$ $0.6$ based on microscopic calculations are $5.39~\rm{MeV}$ and $5.59~\rm{MeV}$, respectively, which both overestimate the experimental value.  In Fig.~\ref{F:MC}, we show the density dependence of the medium correction $\sigma_{\rm NN}^*/\sigma_{\rm NN}^{\rm free}$ at three different $T_{\rm c.m.}$ values for the NN cross section in Eq.~(\ref{E:ECS}) with $\alpha$ $=$ $1.8$, the Rostock cross section with $\alpha$ $=$ $0.6$, and the FU$4$FP$6$ parameterization with the isospin asymmetry $\delta$ set to be $0.21$ as in \isotope[208]{Pb}. The $T_{\rm c.m.}^{\rm F}$ $\approx$ $0.073~\rm{GeV}$ in this figure represents the $T_{\rm c.m.}$ of two nucleons at the Fermi surface of normal nuclear matter density $\rho_{\rm nuc}$. It is seen that both the $\alpha$ $=$ $1.8$ case and the FU$4$FP$6$ parameterization, which can describe the data of the GDR width of \isotope[208]{Pb}, show similar medium reductions, which are very large compared with that from the Rostock cross section.

In the inset of Fig.~\ref{F:gm}, we further show the strength function of the iso-vector excitation of \isotope[208]{Pb} from the LBUU calculation uisng the cross section in Eq.~(\ref{E:ECS}) with $\alpha$ $=$ $1.8$ together with the experimental measurements at RCNP~\cite{TamPRL107}. 
The LBUU result shifted to match the GDR peak energy is also included for comparison.
As can be seen, our calculation nicely reproduces the shape of the experimental strength function, but overestimates the peak energy by about $1.4$ MeV. A better agreement for the peak value could be achieved by varying the mean-field potential or nuclear EOS, which is known to significantly affect the peak energy of nuclear GDR~\cite{TriPRC77}. 
Further tests show that the mean-field potential or nuclear EOS only weakly affects the obtained GDR width.  As to the quantum corrections to the BUU equation~\cite{BonPRL71}, it is not expected to modify our results qualitatively as their effect on the damping of collective motion of heavy nucleus~\cite{KonNPA577} is insignificant compared with that of NN scatterings.


\emph{Conclusions.}
We have used the LH method to solve the BUU transport equation with the binary collisions in the collision term treated via the stochastic approach.
With the use of a sufficiently large number of test particles, the present LBUU method treats the Pauli blocking in the collision term of BUU equation with very high precision and thus significantly increases the accuracy in solving the BUU equation. From the accurately calculated GDR width of \isotope[208]{Pb}, we have found that it depends strongly on the magnitude of the in-medium NN cross section $\sigma_{\rm NN}^*$, and the experimentally measured GDR width of \isotope[208]{Pb} from the $\isotope[208]{Pb}(\vec{p},\vec{p}')$ reaction at RCNP can only be reproduced with a NN cross section that is significantly reduced in nuclear medium. The large medium reduction of $\sigma_{\rm NN}^*$ raises challenges to microscopic calculations based on realistic NN interactions. Also, the effects of such a large medium reduction of $\sigma_{\rm NN}^*$ on the widths of other modes of giant resonances in nuclei and on the dynamics of HICs need to be studied as it may significantly affect the extracted information on the properties of nuclear matter at various densities.


\emph{Acknowledgements.}
We thank Bao-An Li and Jun Su for useful discussions, and Meisen Gao, Jie Pu, Chen Zhong and Ying Zhou for the maintenance of the GPU severs. This work was partially supported by the National Natural Science Foundation of China under Contracts No. $11905302$, No. $11947214$, No. $11890714$, No. $11625521$ and No. $11421505$, the Key Research Program of Frontier Sciences of the CAS under Grant No. QYZDJ-SSW-SLH$002$, the Strategic Priority Research Program of the CAS under Grants No. XDB$16$ and No. XDB$34000000$, the Major State Basic Research Development Program (973 Program) in China under Contract No. $2015$CB$856904$, the US Department of Energy under Contract No. DE-SC$0015266$, and the Welch Foundation under Grant No. A-$1358$.


\end{document}